\documentclass{iopart}
\usepackage{graphicx,amsfonts,amssymb,array,calc,tabularx}
\usepackage{bm}
\usepackage{supertabular}
\usepackage{booktabs}

\begin{document}

\title[GLAST and Lorentz violation]{GLAST and Lorentz violation}
\author{Raphael Lamon}
\address{Institut f\"ur Theoretische Physik, Universit\"at Ulm \\ Albert-Einstein-Allee 11 \\ D-89069 Ulm, Germany}
\ead{raphael.lamon@uni-ulm.de}

\begin{abstract}
We study possible Lorentz violations by means of gamma-ray bursts (GRB) with special focus on the Large Array Telescope (LAT) of GLAST, where we concentrate on models with linear corrections to the speed of light. We simulate bursts with {\it gtobssim} and introduce a Lorentz violating term in the arrival times of the photons. We further perturb these arrival times and energies with a Gaussian distribution corresponding to the time resp. energy resolution of GLAST. We then vary the photon flux in {\it gtobssim} in order to derive a relation between the photon number and the standard deviation of the Lorentz violating term. We conclude with the fact that our maximum likelihood method as first developed in \cite{Lamon:07} is able to make a statement whether Nature breaks the Lorentz symmetry if the number of bursts with known redshifts is of the order of 100. However, the systematic errors caused by unknown mechanisms for photon emission are not considered here despite the fact that these errors should be the main obstacle to detecting Lorentz violations.\index{\footnote{}}
\end{abstract}
\maketitle

\section{Introduction}
The space-time structure as described by general relativity is a classic one in the sense that it is smooth even at very small distances and very large energies. However, there is an agreement that new phenomena caused by quantum gravity (QG) should affect this smoothness at a distance of the order of the Planck length $l_P=\sqrt{\hbar G/c^3}\approx1.6\times10^{-33}$ cm or equivalently the Planck mass $M_P=\sqrt{\hbar c/G}\approx1.2\times10^{19}$ GeV/$c^2$. Despite the fact that we still lack a full quantum theory of gravity we can make some predictions derived from phenomenological quantum gravity. One of these predictions is a distortion of the photon dispersion relation \cite{Pavlopoulos:67,Amelino:97,Amelino:98}
\begin{equation}
 E^2=p^2c^2+\alpha\frac{E^3}{E_P}+\mathcal{O}(E^4/E_P^2),
\end{equation}
where $E$ denotes the photon energy, $p$ its momentum, $\alpha$ a model-dependent dimensionless parameter of order unity and $c$ the speed of light in vacuo. This distortion leads to an energy-dependent velocity of light given by
\begin{equation}\label{ccorr}
 v(E)=c\left(1+\alpha\frac{E}{E_P}\right)+\mathcal{O}\left((E/E_P)^2\right).
\end{equation}
A more general relation with an explicit Lorentz symmetry breaking could be used \cite{Kahniashvili:06}, however there are stringent experimental limits on first order
in $l_P$ modifications to dispersion relations in a lorentz breaking scenario, see e.g. \cite{Fan:07} or the constraint from the Crab \cite{Jacobson:03,Maccione:07}. On the other hand Lorentz invariance could be seen as a consequence of the foamy structure of space-time caused by quantum fluctuations on short times and distance scales \cite{Ellis:92,Ellis:99,Kostelecky:89,Myers:03}. Another approach, called deformed (or doubly) special relativity (DSR) \cite{Amelino:02,Amelino:03,Magueijo:02,Magueijo:02:02,Hossenfelder:07}, introduces a second constant scale (the Planck length $l_P$ or equivalently the Planck energy $E_P$) that is observer independent. This yields a non-linear Casimir operator of DSR which may result in an energy-dependent speed of light, depending on the chosen model. However, a common feature of all DSR models is that the speed of light does not depend on the helicity.

One reason why present experiments have not been able until now to either rule out or confirm a violation of the Lorentz symmetry is the fact that the correction to the velocity of light in Eq.~\eref{ccorr} is suppressed by the Planck energy. However, despite this small effect it was pointed out that one powerful way to look for an energy-dependent velocity of light is given by gamma-ray bursts (GRB) \cite{Amelino:98,Piran:04:02,Jacob:06,Rodriguez:06,Pavlopoulos:05,Scargle:08,Norris:99,Band:04,Bertolami:00,Bertolami:02}. GRBs are the most luminous electromagnetic events occurring in the universe since the Big Bang. They can last from a few milliseconds to minutes, with a typical duration of a couple of seconds \cite{Piran:99,Piran:05}. However, the main problem we face when looking for QG effects in GRB signals is our ignorance of the internal physical processes which are at the origin of the photon emission. Photons with different energies may emanate from different mechanisms, thus further complicating the comparison between events. Despite these difficulties several studies were able to put strong constraints on a Lorentz invariance violation with GRBs \cite{Ellis:99:2,Ellis:03,Ellis:06,Ellis:07}. In a recent work \cite{Albert:07} a preferred range for the linear QG mass scale of $M_{QG}\sim0.4\times10^{18}$ GeV could even be found by studying a flare observed by MAGIC. This result has a sensitivity that probes, for the first time, the Planck mass range.

In this work we shall mainly consider the first-order correction to the speed of light with the conservative bound $\alpha=1$. In other words we assume that a Lorentz symmetry breaking starts to become important only at the Planck scale $M_{\mathrm{QG}}=M_P$, resulting in the following relation for the speed of light:
\begin{equation}\label{speedoflight}
 v(E)=c\left(1\pm\frac{E}{M_Pc^2}\right).
\end{equation}
The dependence of the speed of light on the energy as described by Eq.~\eref{speedoflight} causes a difference in the arrival times between two photons emitted at the same time, but with different energies. Assuming a flat universe described by the $\Lambda$CDM model (for different models see \cite{Biesiada:07}), the integrated time delay between two photons with an energy difference $\Delta E$ is given by \cite{Jacob:08}
\begin{equation}
 \Delta t=\pm H_0^{-1}\frac{\Delta E}{M_P c^2}\int_0^z\frac{1+z'}{\sqrt{\Omega_{\Lambda}+\Omega_m(1+z')^3}}dz',
\end{equation}
where $\Omega_m=0.27$, $\Omega_{\Lambda}=0.73$ and $H_0=71$ km $\mathrm{s}^{-1}$ $\mathrm{Mpc}^{-1}$. At the end of the paper we will briefly discuss the case of a quadratic correction where the integrated time delay is given by
\begin{equation}\label{quadratic}
 \Delta t=\pm \frac{3}{2}H_0^{-1}\left(\frac{\Delta E}{M_P c^2}\right)^2\int_0^z\frac{(1+z')^2}{\sqrt{\Omega_{\Lambda}+\Omega_m(1+z')^3}}dz'.
\end{equation}

The paper is organized as follows. In \sref{sec:GLAST} we describe the relevant properties of GLAST and in \sref{sec:analysis} and \sref{sec:spectrum} our method to create a photon list with arrival times and energies. \Sref{sec:results} is devoted to explaining our results, \sref{sec:quadratic} to giving an estimate of GLAST to quadratic corrections to the speed of light and \sref{sec:non-FRED} to studying non-FRED distributions. We then conclude with \sref{sec:Conclusion}.

\section{GLAST}\label{sec:GLAST}
\begin{figure}[!ht]
 \begin{center}
  \includegraphics[width=6cm]{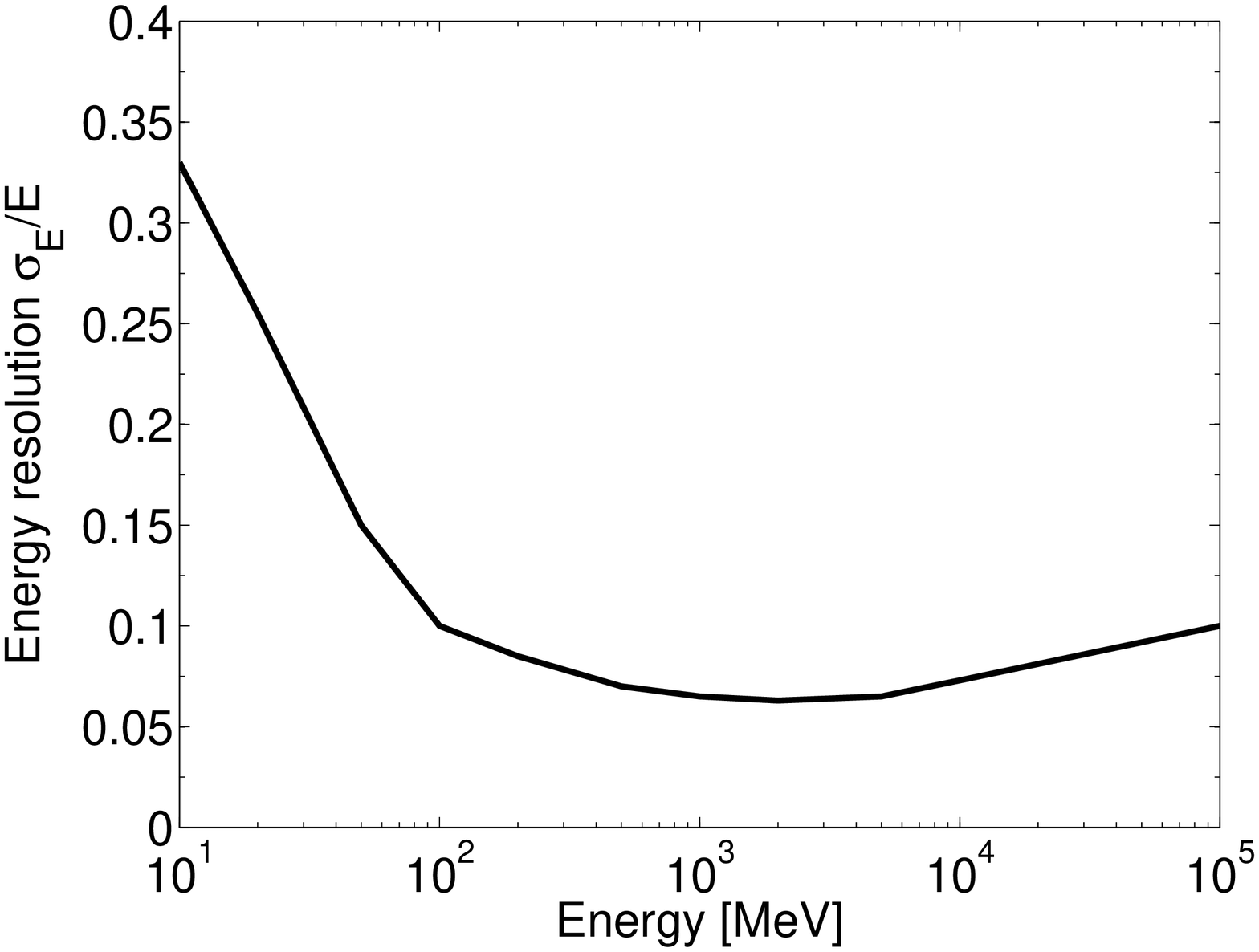}
\caption{Energy resolution as a function of the energy for the LAT \cite{LATperf}. The energy uncertainty at 30 MeV is about 17\% before going down to a couple of percent in the GeV range. Most of the photons LAT detects have an energy around $\sim100$ MeV with an uncertainty of about 10\%.}\label{fig:energyres}
 \end{center}
\end{figure}

The Gamma Ray Large Area Space Telescope (GLAST) is a space-based gamma-ray telescope designed to explore the high-energy universe. It includes two instruments: the Large Array Telescope (LAT), which is an imaging gamma-ray detector which detects photons with energy from about 30 MeV to 300 GeV, and the Gamma-ray Burst Monitor (GBM) that consists of 14 scintillation detectors which detect photons with an energy between 8 keV and 30 MeV. The LAT has a very large field of view that allows it to see about 20\% of the sky at any time. Despite the fact that it will cover the entire sky every three hours we shall assume that the instrument response changes on timescales longer than a typical burst duration. On the other hand, GLAST can be pointed as needed when a bright GRB is detected by either LAT or GBM so that it will detect around 200 GRBs each year. The energy resolution ranges from 20\% at 30 MeV to about 7\% at 1 GeV, as can be seen in \fref{fig:energyres} \cite{LATperf}. The time resolution of an event should be around 10 $\mu$s with a dead time shorter than $100 \mu$s. In summary, LAT will have superior area, angular resolution, field of view, time resolution and deadtime. This will at least provide an advance of a factor 30 in sensitivity compared to previous missions.

The GLAST Science Support Center (GSSC) provides analysis tools freely available to the scientific community on their homepage \cite{GSSC}. As we are interested in simulating the detection of GRBs by GLAST we mainly used the tool called {\it gtobssim} which is a software that generates photon events from astrophysical sources with the instrument response functions of GLAST. Because we will only study the measurement of possible Lorentz violation we assumed that LAT pointed in the same direction as the burst. Further information on the effects of QG on LAT GRBs can be found in e.g. \cite{Norris:99,Band:04}.

\section{Creation of a photon list}

\subsection{Analysis method}\label{sec:analysis}
\begin{figure}[!ht]
 \begin{center}
  \includegraphics[width=6cm]{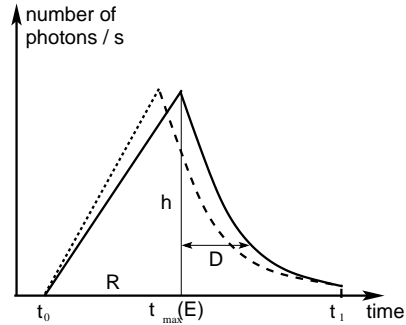}
\caption{Sketch of a typical light curve of a GRB for a given energy interval. The curve is parameterized by four parameters: $ R $ the duration of the rise, $ h $ the height above the background, $ D $ the decay time for $\exp(-t/D)$ and  $\kappa$ describes the magnitude of the dependence on the energy of the distribution $f$, $ t_{\mathrm{max}}=P+\kappa\cdot E$, where $P$ is the time when the intensity reaches a maximum and $E$ is the photon energy. The dashed line shows a distribution for another energy interval that is shifted by an amount of $\Delta t=\kappa\cdot\Delta E$ sketching the shift in time due to quantum gravitational effects.}\label{fig:FRED}
 \end{center}
\end{figure}

We shall use the same method we used to study INTEGRAL GRBs by modeling bursts with a Fast Raise and Exponential Decay (FRED) distribution \cite{Lamon:07}. We parameterize this distribution with four parameters: $f=f(t_i,E_i;P,R,D,\kappa,h)$ as shown in \fref{fig:FRED}. We suppose that a photon $i$ came from the probability density function $f$ at time $t_i$ with energy $E_i$. We use the method of maximum likelihood which consists in finding the set of values $\hat P$, $\hat R$, $\hat D$, $\hat{\kappa}$ and $\hat h$ that maximizes the joint probability distribution of all data, given by
\begin{equation}\label{maxlikelihood}
\mathcal{F}(P,R,D,\kappa,h) =\prod_i f(t_i,E_i;P,R,D,\kappa,h)
\end{equation}
together with the constraint
\begin{equation}\label{constraint}
\int_{t_0}^{t_1} dt' \, f(t',E_i;P,R,D,\kappa,h) = 1,
\end{equation}
where $\mathcal{F}$ is the likelihood function and the integral runs between $t_0$ and $t_1$ as shown in \fref{fig:FRED}. The specifics of this distribution does not play a significant role in our study as we are only interested in seeking possible violations of the Lorentz symmetry. In order to further improve our model we also simulated an isotropic background with a spectrum following an exponential decay with exponent 2.1. However, this background does not play a significant role because of the following two reasons. The first one is that LAT shall only be able to detect photons with energies above $\sim 30$ MeV, where such events are rare. The second one is the fact that we are only interested in GRBs, e.g. events of short duration.

\begin{figure}[!ht]
 \begin{center}
 \includegraphics[width=5.5cm]{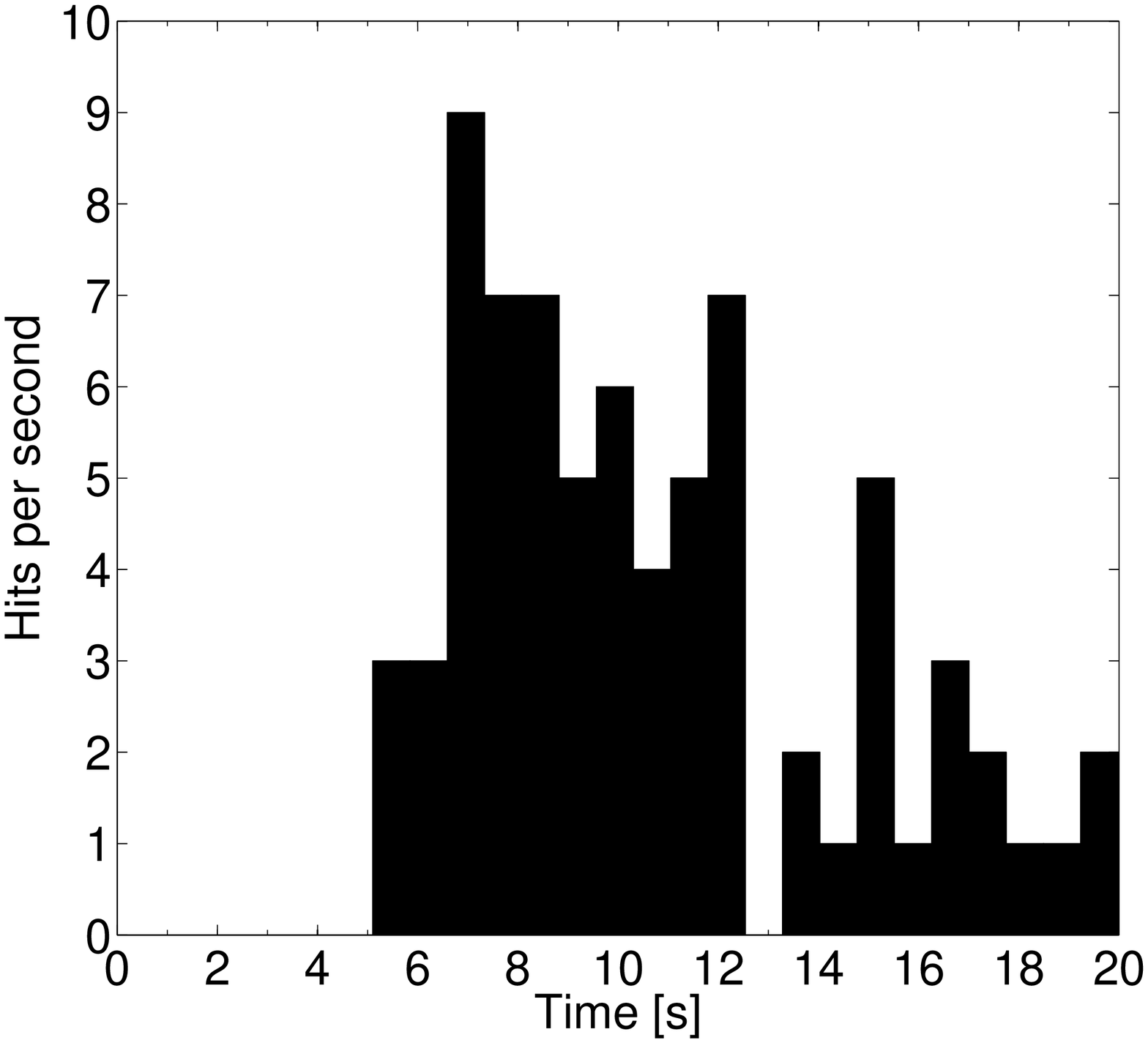}
  \includegraphics[width=5.5cm]{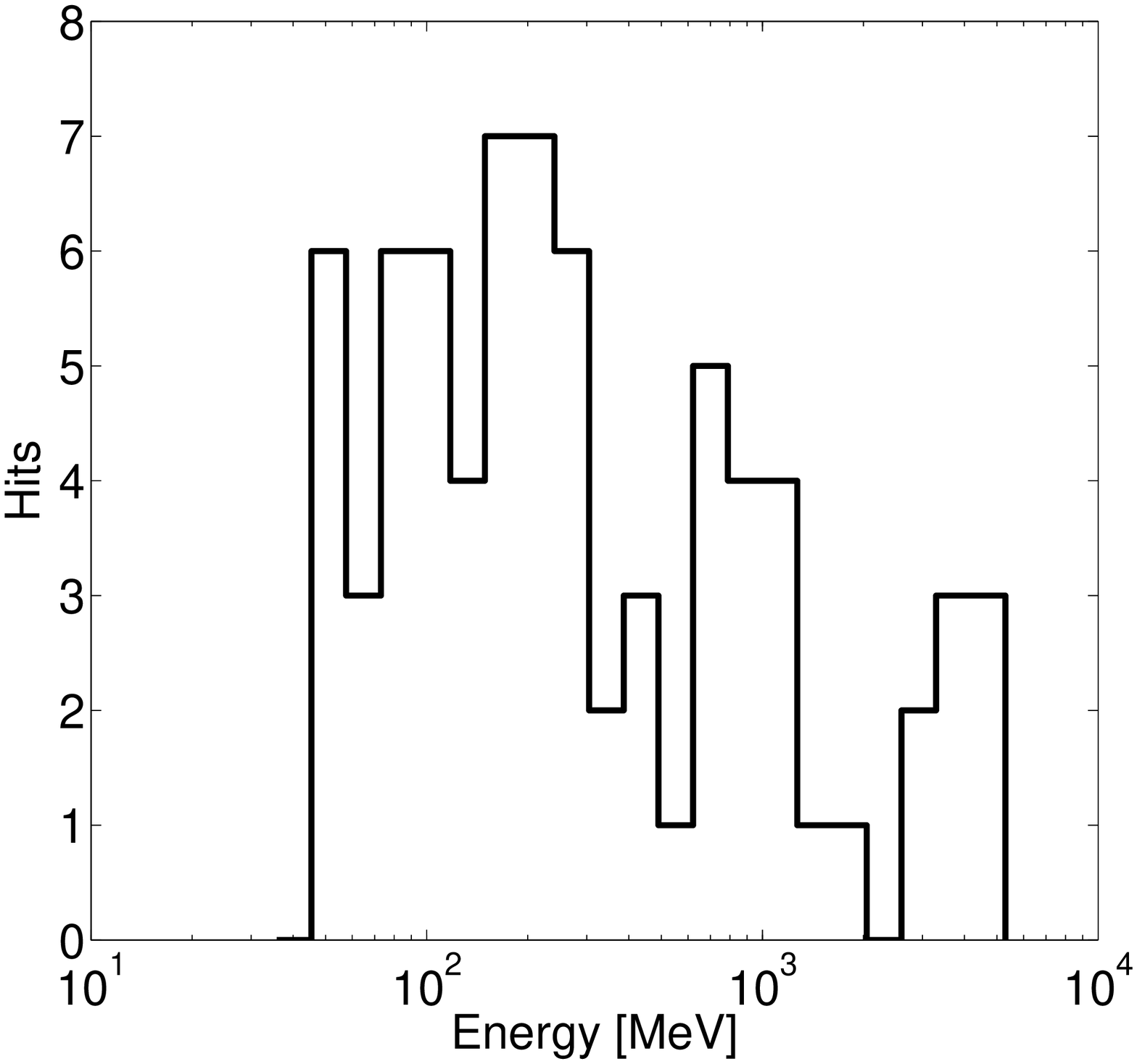}
\caption{{\it Left panel}: Example of a simulated LAT GRB with {\it gtobssim} with a total photon number of 74. The burst starts at $t_0=5$ s and has a decay time of the order of 10 seconds. {\it Right panel}: Spectrum of the same GRB as simulated by {\it gtobssim} for LAT. The first detected photons have an energy of roughly 30 MeV. We choose the high energy exponent $\beta=2$ (see Eq.~\eref{BAND}) and a background with exponent $\gamma=2.1$}\label{fig:GRB}
\end{center}
\end{figure}

\subsection{Spectrum of the GRBs}\label{sec:spectrum}
As described in the previous section we also need to simulate the energy of the photons. Normally a typical energy distribution of GRBs follows the pattern of the so-called Band function \cite{Band:93} given by the following equation:
\begin{eqnarray}\label{BAND}
N_E(E)&=&A\left( \frac{E}{100\;\mathrm{keV}}\right) ^{\alpha}\exp\left( -\frac{E}{E_0}\right) ,\nonumber\\
&& \quad\quad\quad\quad\quad\quad\quad\quad\quad\quad\quad(\alpha-\beta)E_0\geq E,\nonumber\\
&=&A\left[ \frac{(\alpha-\beta)E_0}{100\;\mathrm{keV}}\right] ^{\alpha-\beta}\left( \frac{E}{100\;\mathrm{keV}}\right) ^{\beta}\exp(\beta-\alpha),\nonumber\\
&& \quad\quad\quad\quad\quad\quad\quad\quad\quad\quad\quad(\alpha-\beta)E_0\leq E,
\end{eqnarray}
where $\alpha$ is the low-energy exponent, $\beta$ the high-energy one and $E_0$ the break energy. As LAT starts measuring at 30 MeV and the break energy is around 500 keV we shall only be interested in the high-energy behavior of the Band function. However, LAT will open a new window on the spectrum where little is known, therefore there is no certainty whether the Band function is still valid throughout the energy range of LAT.

\begin{figure}[!ht]
 \begin{center}
 \includegraphics[width=12cm]{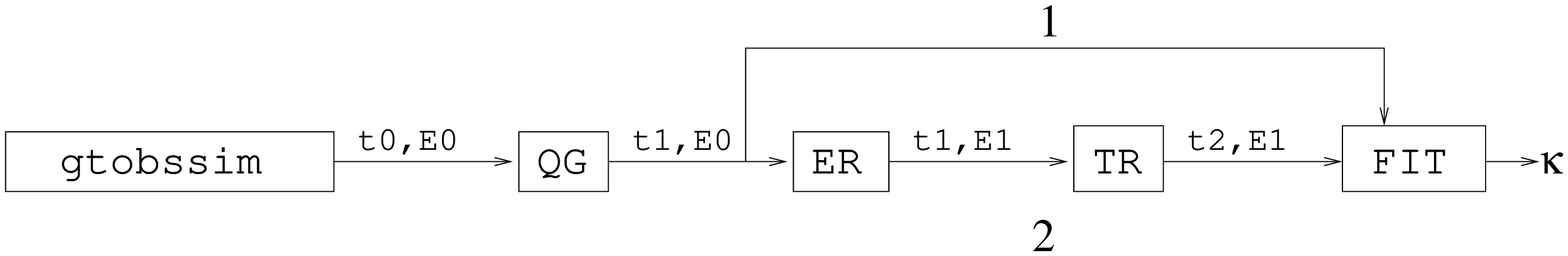}
\caption{Setup of the simulation process. The software {\it gtobssim} gives as output the arrival time $t_0$ and the energy $E_0$ of a photon. The arrival time is then corrected by means of the relation~\eref{intrk} from $t_0$ to $t_1$ (box QG). The simpler case with only QG effects  is shown by the arrow number 1. In the second case, the energy of the photon was perturbed from $E_0$ to $E_1$ with a Gaussian distribution with standard deviation according to \fref{fig:energyres} (box ER). Furthermore, the arrival time was perturbed a second time from $t_1$ to $t_2$ with a Gaussian distribution with standard deviation of 10 $\mu$s corresponding to the time resolution of LAT (box TR). The photon with arrival time $t_2$ and energy $E_1$ is then given to the fitter (box FIT) to get a value for the QG corresponding to $\kappa$. }\label{fig:process}
\end{center}
\end{figure}

In order to make the simulations as realistic as possible we used {\it gtobssim}. As described in the previous section we simulated a GRB with photons following a FRED distribution together with an isotropic background. Since we are only interested in the detection of a possible Lorentz violation we used the same FRED distribution for all GRBs, i.e. with a raise time of the order of a second and a decay time around 10 seconds (see \fref{fig:GRB}). In order to get a feeling of the uncertainty we varied the flux of the burst and only selected the bursts which got about the same number of hits in the detector. We then introduced a Lorentz violating term with the following relation (see \fref{fig:process}):
\begin{equation}\label{intrk}
 \Delta t=\kappa\times E_{\gamma},
\end{equation}
where $E_{\gamma}$ is the energy of the detected photon in MeV and $\Delta t$ is the time delay in seconds caused by quantum gravity given by the relation $t_1=t_0+\Delta t$. The parameter $\kappa$ describes the effect of a Lorentz violation and is given by the following relation:
\begin{equation}
 \kappa=\frac{H_0^{-1}}{M_{\mathrm{QG}}c^2}\int_0^z\frac{1+z'}{\sqrt{\Omega_{\Lambda}+\Omega_m(1+z')^3}}dz'=:\frac{H_0^{-1}}{M_{\mathrm{QG}}c^2}I(z),
\end{equation}
where $M_{\mathrm{QG}}$ is the mass scale where the Lorentz symmerty breaks down. Henceforth we shall take the most conservative scale and set $M_{\mathrm{QG}}=M_P\approx1.2\times10^{19}$ GeV. 

\begin{figure}[!ht]
 \begin{center}
 \includegraphics[width=6cm]{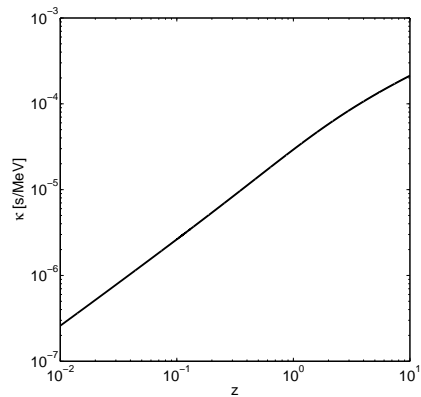}
\caption{The Lorentz violation parameter $\kappa$ as a function of the redshift. As we shall see below, a value of $\kappa=10^{-4}$ s/MeV could even be measured without requiring a bright burst (see \tref{table:results}).} \label{fig:kappa}
\end{center}
\end{figure}

As can be seen in \fref{fig:kappa} typical values for $\kappa$ lie between $10^{-5}$ and $10^{-4}$ s/MeV. In the Monte Carlo simulations we used a value of $\kappa=4\times10^{-5}$ corresponding to a redshift of $z=1$.

We studied two models: one simple one with only the time delays caused by quantum gravity and a more realistic one where we perturbed the energy and the arrival time according to the scheme shown in \fref{fig:process} in order to take into account the energy and time resolution of the LAT. For a given event at time $t_0$ with energy $E_0$ obtained with {\it gtobssim} we read the energy uncertainty $\sigma_{E_0}/E_0$ from \fref{fig:energyres} and perturbed this energy with a Gaussian with maximum at $E_0$ and standard deviation $\sigma_{E_0}$. The next step is to take the corrected time $t_1$ and perturb it with a Gaussian with maximum at $t=0$ and standard deviation $\sigma_t=10$ $\mu$s so that we get the perturbed arrival time $t_2$.

\subsection{Results}\label{sec:results}
In the previous section we explained how we constructed a burst with photons and perturbed their energies and arrival times to account for Lorentz violation and finite energy and time resolution of LAT. The question is now whether it is possible to get back the value of $\kappa$ parameterizing the Lorentz violating term despite the perturbation of both the energy and arrival time. In \cite{Lamon:07} we found an exponential dependence between the standard deviation $\sigma_{\kappa}$ and the number of detected photons $N$ with an exponent of -0.617. However, we only perturbed the arrival time and not the energy.

The final step is to search for $\bar{\kappa}$ which minimizes the likelihood \eref{maxlikelihood}. For a single burst we scanned through a large range of $\kappa$ in order to find the global (and not just a local) minimum of the likelihood. We performed 100 Monte Carlo simulations for different luminosities of the bust, i.e. different numbers of total events $N$ (see \tref{table:results}).

 \begin{table}[!ht]
\begin{center}
 \begin{tabular}{llllll}
& & \multicolumn{2}{c}{unperturbed (1.)} & \multicolumn{2}{c}{perturbed (2.)}\\
N & flux [s$^{-1}$m$^{-2}$] & $\bar{\kappa}$ [s/MeV] & $\sigma_{\kappa}$ [s/MeV] & $\bar{\kappa}$ [s/MeV] & $\sigma_{\kappa}$ [s/MeV]\\
\midrule
20 & 10 & $1.2\times10^{-4}$ & $2.0\times10^{-3}$ & $3.9\times10^{-4}$ & $1.7\times10^{-3}$\\
50 & 28 & $9.5\times10^{-5}$ & $9.5\times10^{-4}$ & $6.4\times10^{-5}$ & $7.8\times10^{-4}$\\
75 & 46 & $8.5\times10^{-5}$ & $5.3\times10^{-4}$ & $1.4\times10^{-4}$ & $2.4\times10^{-4}$ \\
100 & 60 & $6.5\times10^{-5}$ & $4.3\times10^{-4}$ & $8.4\times10^{-5}$ & $3.7\times10^{-4}$ \\
150 & 96 & $4.2\times10^{-5}$ & $2.2\times10^{-4}$ & $4.6\times10^{-5}$ & $3.3\times10^{-4}$ \\
200 & 128  & $5.3\times10^{-6}$ & $2.1\times10^{-4}$ & $3.7\times10^{-5}$ & $2.4\times10^{-4}$\\
500 & 330 & $2.9\times10^{-5}$ & $6.6\times10^{-5}$ & $1.4\times10^{-5}$ & $6.6\times10^{-5}$ \\
1000 & 660 & $2.3\times10^{-5}$ & $5.1\times10^{-5}$ & $2.9\times10^{-5}$ & $4.3\times10^{-5}$ \\
\end{tabular}
\end{center}
\caption{Results of 100 Monte Carlo simulations. The first column shows the number of photons detected by LAT, the second one the approximate flux in $\mathrm{m}^{-2}\mathrm{s}^{-1}$. The third column shows the mean value of $\bar{\kappa}$ for the unperturbed system (1.) and the fourth one its standard deviation. The last two columns are the same as the third and fourth ones, except for the fact that the arrival times and energies were perturbed with a Gaussian distribution (see \sref{sec:spectrum}).}\label{table:results}
\end{table}


\begin{figure}[!ht]
 \begin{center}
 \includegraphics[width=6cm]{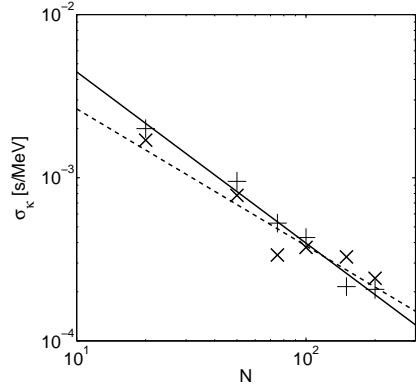}
\caption{Comparison between the standard deviations $\sigma_{\kappa}$ of the unperturbed and perturbed system. The vertical crosses show the values for the unperturbed system together with the solid line as its fit. The inclined crosses show the values for the system with perturbed arrival times and energies.}\label{fig:sigma}
\end{center}
\end{figure}

\Fref{fig:sigma} shows a comparison between the standard deviation of the perturbed and unperturbed system. The fit of the unperturbed system is given by
\begin{equation}
 \sigma_{\kappa}=(2.5\times10^{-2})\cdot N^{-0.93}
\end{equation}
and the fit of the perturbed system by
\begin{equation}\label{sigmapert}
 \sigma_{\kappa}=(6.0\times10^{-2})\cdot N^{-1.07}.
\end{equation}

\begin{figure}[!ht]
 \begin{center}
 \includegraphics[width=7cm]{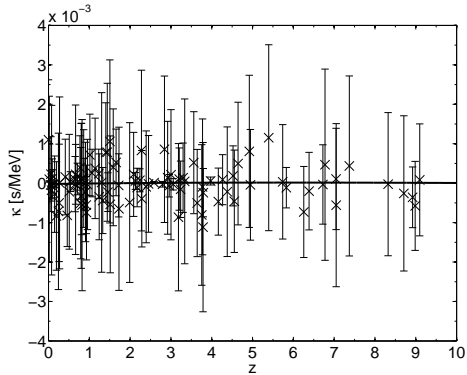}
\caption{Simulation of 100 GRBs where $\kappa$ has been perturbed with a Gaussian according to the relation~\eref{sigmapert}. The error bars show the 1$\sigma$ deviation given by Eq.~\eref{sigmapert}. The distribution of the GRBs and the luminosity has been computed using results from \cite{Omodei:07,Omodei:06} where a histogram of the number of GRBs as a function of either the redshift or the luminosity is given. The solid line shows the fit given by Eq.~\eref{fit}. Note that no systematic errors have been incorporated.}\label{fig:kappasim}
\end{center}
\end{figure}

The results described above allows us to simulate a full set of GRBs with different redshifts $z$ and time shifts $\kappa$. \Fref{fig:kappasim} shows a set of 100 GRBs with redshifts between 0 and 10, where we used the results from \cite{Omodei:07} and \cite{Omodei:06} in order to get a rough estimate of the expected number of GRBs detected by GLAST as a function of the number of detected photons and the redshift. Each value for $\kappa$ has been perturbed with a Gaussian with a standard deviation given by Eq.~\eref{sigmapert}. Considering a linear approximation to quantum gravitational effects we have the relation \cite{Ellis:03,Ellis:06}:
\begin{equation}\label{fitth}
 \kappa=aI(z)+b(1+z),
\end{equation}
where $a$ and $b$ are fitted coefficients. The constant $b$ parameterizes time lags in the frame of the source caused by unknown internal processes of the GRBs, $a$ describes the expected Lorentz violating effects through
\begin{equation}\label{ath}
 a_{\mathrm{th}}=\frac{H_0^{-1}}{M_Pc^2}\approx 2.56\times 10^{-5}\;\;\mathrm{s}/\mathrm{MeV}.
\end{equation}

A fit of the 100 GRBs shown in \fref{fig:kappasim} gives
\begin{equation}\label{fit}
 \kappa=(2.6\pm3.1)\times10^{-5}\cdot I(z)-(1.9\pm2.5)\times10^{-5}\cdot (1+z).
\end{equation}
We see that the Lorentz violating term is well compatible with the theoretical value $a_{\mathrm{th}}$ while the second term is less than one $\sigma$ off the input values $b_{\mathrm{th}}=0$ (see \tref{table:NGRB}). Considering the fact that LAT should see around 500 GRBs each year it might be tantalizing to conclude that only after a couple of months the question whether the Lorentz symmetry is broken might be answered. However, we would like to stress the fact that we did not consider systematic errors caused by the lack of knowledge of the internal processes leading to photon emissions. As these systematic errors were the main problem past works \cite{Lamon:07,Ellis:03,Ellis:06,Ellis:07} had to deal with, this oversimplified analysis must be taken with precaution. Moreover, a known redshift of the bursts is also needed, thus reducing the number of usable bursts.

\begin{table}[!ht]
\begin{center}
 \begin{tabular}{lll}
  $N$ & $a$ [s/MeV] & $b$ [s/MeV]\\
\midrule
10 & $(1.7\pm5.6)\times10^{-5}$ & $(-1.8\pm3.3)\times10^{-5}$\\
20 & $(2.6\pm3.6)\times10^{-5}$ & $(-1.8\pm2.8)\times10^{-5}$\\
50 & $(2.6\pm3.2)\times10^{-5}$ & $(-1.9\pm2.6)\times10^{-5}$\\
100 & $(2.6\pm3.1)\times10^{-5}$ & $(-1.9\pm2.5)\times10^{-5}$
 \end{tabular}
\end{center}
\caption{Results of the fit~\eref{fitth} for a variable number of bursts. The first column shows the number of bursts considered for the fit, the second one shows the value of $a$ describing the Lorentz violating effects  and the third one the value of $b$ parameterizing the time lags caused by unknown internal processes. The theoretical value of $a$ is given by Eq.~\eref{ath}. The error on $a$ and $b$ only reduces considerably between $N=10$ and $N=20$.}\label{table:NGRB}
\end{table}

\subsection{Quadradic corrections}\label{sec:quadratic}
In this subsection we are concerned with quadratic corrections given by Eq.~\eref{quadratic}. Such corrections may be interesting in view of the fact that previous works have already put stringent constraints on linear corrections to the speed of light \cite{Jacobson:03,Fan:07,Ellis:07,Albert:07}. To get a rough estimate of the sensitivity of GLAST to quadratic corrections we must first get a bound on the time difference that should be detectable by GLAST. With Eq.~\eref{intrk} we see that for $\sigma_{\kappa}\sim 5\times 10^{-4}$ s/MeV (see \tref{table:results}) and $\Delta E\sim 10^3$ MeV we get a bound on the time difference of $\sigma_{\Delta t}\sim 5\times 10^{-1}$ s. Solving Eq.~\eref{quadratic} for the mass scale and inserting these results we get a mass scale of
\begin{equation}
 M_{\mathrm{quadratic}}\gtrsim2\times10^{9}\;\mathrm{GeV}.
\end{equation}
for $z=1$ up to which GLAST should be sensitive for a \textit{single} burst. This bound may probably be raised by a couple of orders of magnitude with better statistics. However, we do not think that GLAST will be able to detect a quadratic correction if the QG scale is at the Planck scale.

The above estimate may seem very crude. We therefore checked it with results obtained in the literature and found a good agreement between this estimate and a more rigorous statistical analysis. For example, taking $M_L>7\times10^{15}$ GeV obtained in Eq.~(36) in \cite{Ellis:03} together with $\Delta E\sim100$ keV and $\sigma_{\Delta t}\sim 5\times10^{-3}$ s for BATSE we get a bound on the quadratic correction of $M_Q\gtrsim 0.8\times 10^6$ GeV, which corresponds more or less to the result $M_Q>3\times10^6$ GeV obtained in \cite{Ellis:03} with a rigorous analysis.

\subsection{Other distributions}\label{sec:non-FRED}
\begin{figure}[!ht]
 \begin{center}
 \includegraphics[width=5.5cm]{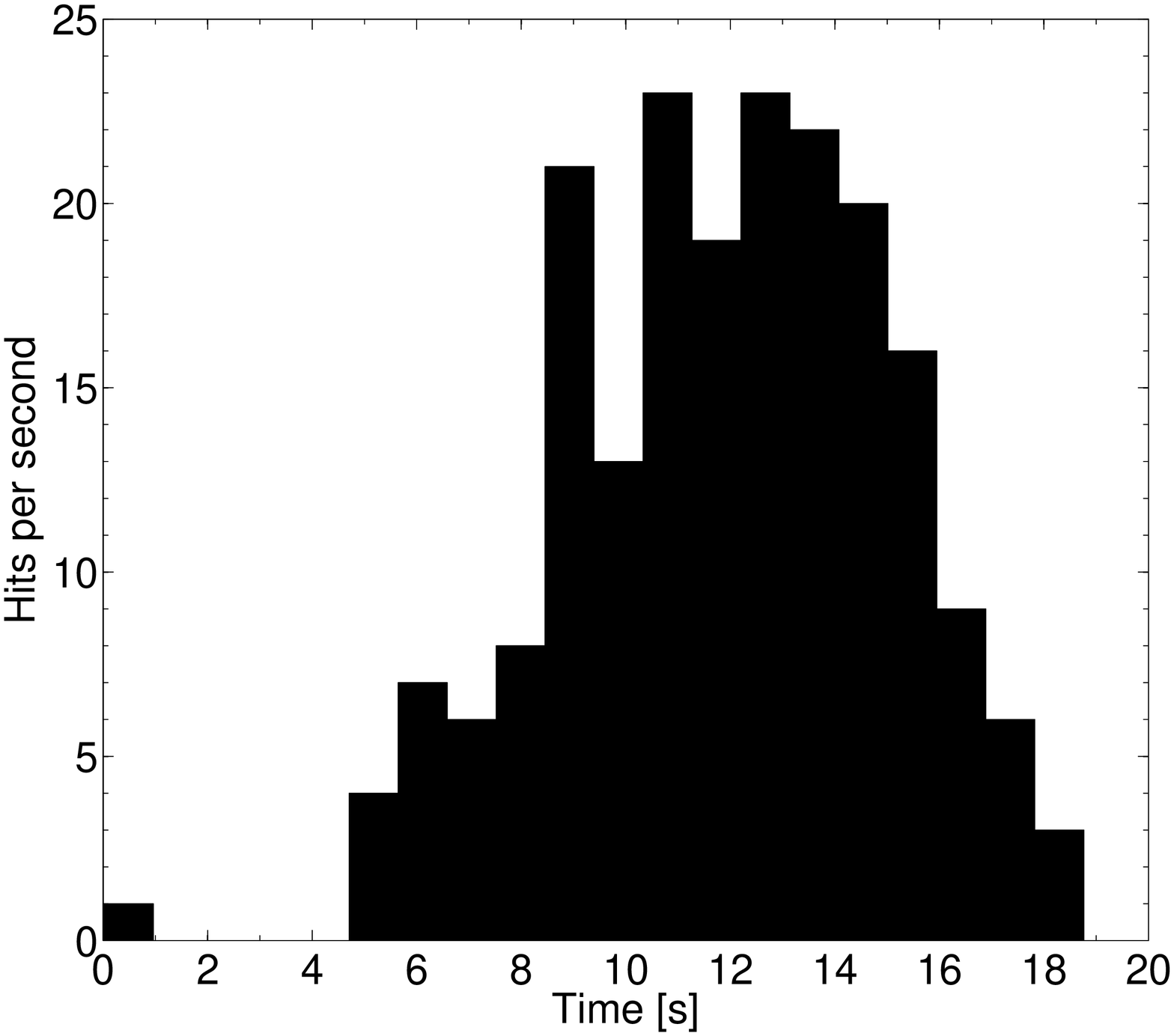}\includegraphics[width=5.5cm]{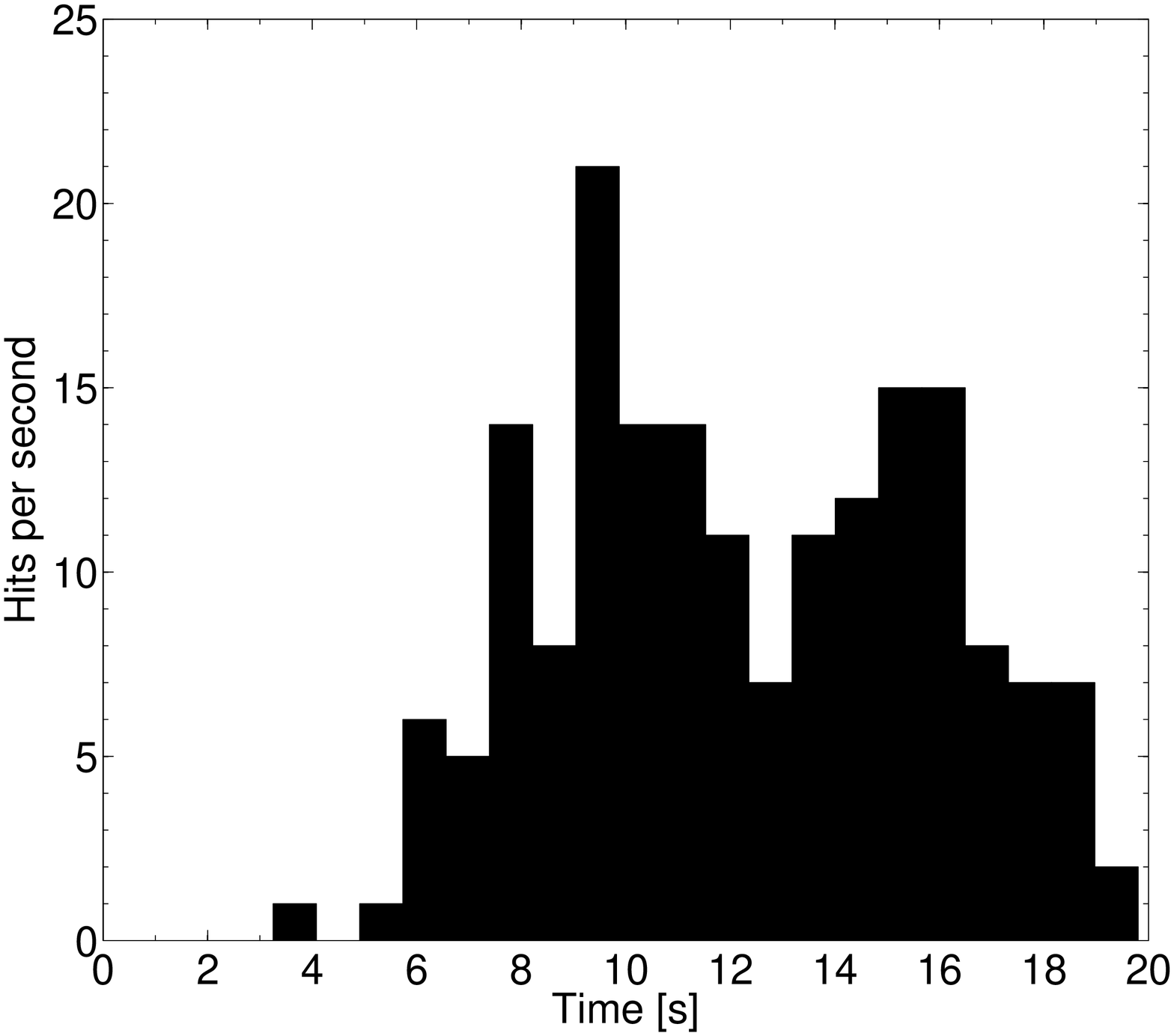}
\caption{\textit{Left panel}: Example of a simulated LAT GRB with a linear raise and linear decay. \textit{Right panel}: Example of a simulated LAT GRB with a double-peak shape.}\label{fig:non-FRED}
\end{center}
\end{figure}

Until now we only considered FRED distributions for GRBs. However, despite the fact that these distributions may describe a large number of GRBs, we also have to consider other burst shapes in order to get an estimate for the validity of our likelihood method with non-FRED distributions. We studied two other shapes, a linearly raising and falling distribution and a double-peak distribution (see \fref{fig:non-FRED}). We then applied the likelihood method to these peaks with 200 photons and compared the results with \tref{table:results} for $N=200$. For the linear peak (see left panel in \fref{fig:non-FRED}) we found a value $\bar{\kappa}=1.7\times 10^{-6}$ s/MeV and a standard deviation of $\sigma_{\kappa}=1.6\times10^{-4}$ s/MeV. Comparing this value with the perturbed system for $N=200$ (see \tref{table:results}) we see that the likelihood method gives a slightly better results for the linear peak. We also studied the double peak (see fight panel in \fref{fig:non-FRED}) and found a value of $\bar{\kappa}=2.0\times10^{-6}$ s/MeV with a standard deviation of $\sigma_{\kappa}=1.6\times 10^{-4}$. Thus the results of both cases lead to the conclusion that our likelihood method is also able to deal with non-FRED distributions.

We fixed the duration time of the simulated GRBs to about 20 seconds, which seems quite arbitrary. However, we expect the likelihood to get better as the duration of the burst gets shorter (as long as the total number of photons remains constant). The reason is that, while $\kappa$ is independent of the duration, the ratios $\kappa/R$ and $\kappa/D$ (see \fref{fig:FRED}) are inverse proportional to $t_1-t_0$. We checked this claim for bursts with a duration time of about 6 seconds, 200 photons and FRED-distributed. We found a value $\kappa=2.6\times10^{-5}$ s/MeV with a standard deviation of $\sigma_{\kappa}=3.7\times10^{-5}$ s/MeV. Comparing this value with \tref{table:results} we see that the likelihood is able to recover the Lorentz violating term with a precision of about one order of magnitude better. On the other hand, the convergence of the likelihood decreases with the duration time of the bursts, thus narrowing the study of short bursts to the brighter ones.

\section{Conclusion}\label{sec:Conclusion}
As can be seen from \tref{table:results} values of the Lorentz violation parameter $\kappa$ of the order of $\sim10^{-5}$ s/MeV could theoretically be measurable. In \cite{Omodei:06,Omodei:07,Omodei:07:02} the BATSE catalog was used to build up the statistics for LAT. The authors made the assumption of extrapolating the BATSE observations to LAT energies in order to get a rough estimate on the number of detected bursts each year as a function of the number of photons per burst and the energy threshold (see Table~1 in \cite{Omodei:07}). If this assumption is correct we may get a couple of bursts each year that could answer the question whether the Lorentz symmetry is violated. Nevertheless, as explained above, the systematics should considerably wash out the signal from a Lorentz violating term so that a lot of bursts with known redshifts should be needed to get a tight bound on a Lorentz violating term. Even without systematics a number of bursts of the order of 100 still yields an uncertainty of the order of the expected Planck corrections, as can be seen from Eq.~\eref{fit}.

In this work we had to make a couple of simplifications. For example we had to assume that the BAND function is still valid at LAT energies. In other words, GLAST will be able to measure the high-energy index at energies out of reach of past and present experiments. Furthermore, we studied the effect of a linear correction to the speed of light and assumed a simultaneous emission of the photons of different energies from the source. This last assumption is the most problematic one because in \cite{Lamon:07} we obtained results for $\kappa$ with even different signs for the same burst (see also e.g. \cite{Ellis:03,Ellis:06,Rodriguez:06:02} where the same kind of problems arises). Therefore, the unknown emission mechanisms for photon emission should be the major problem in seeking Lorentz violation. This is the reason why either a lot of bursts or a better understanding of the burst mechanism is needed before we can settle down the question whether the Lorentz symmetry is violated in Nature.

\section{Acknowledgment}
I would like to thank Nicolas Produit, Frank Steiner and Eric Str\"ang for correcting this manuscript.

\section*{References}

\end{document}